\documentstyle[12pt]{article}
\newcommand{\be}{\begin{equation}}
\newcommand{\ee}{\end{equation}}
\newcommand{\ba}{\begin{eqnarray}}
\newcommand{\ea}{\end{eqnarray}}
\newcommand{\bi}[1]{\bibitem{#1}}
\newcommand{\fr}[2]{\frac{#1}{#2}}
\newcommand{\non}{\nonumber}
\newcommand{\ar}{\mbox{$\rightarrow$}}

\newcommand{\al}{\mbox{$\alpha$}}
\newcommand{\ka}{\mbox{$\kappa$}}
\newcommand{\val}{\mbox{$\vec{\alpha}$}}
\newcommand{\om}{\mbox{$\omega$}}
\newcommand{\Z}{\mbox{$Z\alpha$}}
\newcommand{\p}{\mbox{$\vec{p}$}}
\newcommand{\pp}{\mbox{$\vec{p}\,'$}}
\newcommand{\rp}{\mbox{$\vec{r}\,'$}}
\newcommand{\k}{\mbox{$\vec{k}$}}
\newcommand{\kp}{\mbox{$\vec{k'}$}}
\newcommand{\r}{\mbox{$\vec{r}$}}

\newcommand{\dE}{\mbox{$\Delta E$}}
\newcommand{\D}{\mbox{$\vec{D}$}}
\newcommand{\K}{\mbox{$\cal{K}$}}
\newcommand{\C}{\mbox{$\cal{C}$}}
\newcommand{\M}{\mbox{$\cal{M}$}}
\newcommand{\Sg}{\mbox{$\cal{S}$}}
\newcommand{\G}{\mbox{$\cal{G}$}}
\newcommand{\lb}{\left (}
\newcommand{\rb}{\right )}
\newcommand{\la}{\left\langle}
\newcommand{\ra}{\right\rangle}
\def\Res{\mathop{\rm Res}}

\begin{document}
\pagestyle{empty}
\vspace{1.0cm}

\begin{center}
{\Large \bf Recoil Correction to Hydrogen Energy Levels:\\ A Revision }

\bigskip

{\bf A.S. Yelkhovsky}\\
Budker Institute of Nuclear Physics,\\
and \\
Physics Department, Novosibirsk University, \\
630090 Novosibirsk, Russia
\end{center}

\bigskip

\begin{abstract}
Recent calculations of the order $(Z\alpha)^4\frac{m}{M}$Ry pure recoil
correction to hydrogen energy levels are critically revised. The origins of
errors made in the previous works are elucidated. In the framework of a
successive approach, we obtain the new result for the correction to $S$
levels. It amounts to $-16.4$ kHz in the ground state and  $-1.9$ kHz in the
$2S$ state.
\end{abstract}

\newpage

\pagestyle{plain}
\pagenumbering{arabic}

\section{Introduction}

The correction to $S$ levels of hydrogen atom, that is first-order in $m/M$
and fourth-order in \Z, has become recently a point of controversy.
Initially, this correction was calculated in Ref.\cite{PG}. Then, a
different result for the same correction was obtained in Ref.\cite{I}. While
in both papers it was employed the same (exact in \Z) starting expression
for the pure recoil correction, the methods of calculation, and in
particular the regularization schemes used were rather different. To resolve
the discrepancy between two results, in Ref.\cite{EG} an attempt was
undertaken to prove the correctness of the earlier result of Ref.\cite{PG}
applying the method of calculation used by the present author in
Ref.\cite{I}. An extra contribution due to the peculiarities of the
regularization procedure was found by the authors of Ref.\cite{EG}, which
exactly compensated the difference of the Ref.\cite{I} result from that of
Ref.\cite{PG}. This finding has led the authors of Ref.\cite{EG} to
conclusion "that discrepancies between the different results for the
correction of order $(\Z)^6(m/M)$ to the energy levels of the hydrogenlike
ions are resolved and the correction of this order is now firmly
established".

Taking criticism of the Ref.\cite{EG} as completely valid, we nevertheless
cannot agree with the conclusion cited above. The point is that the authors
of Ref.\cite{EG} emphasizing an importance of an explicit regularization of
divergent expressions, pay no attention to an accurate matching of
regularized contributions.

In fact, one usually starts from an exact expression which can be easily
checked to have a finite value. Then one has to use different approximations
to
handle this expression at different scales. In this way some auxiliary
parameter(s) are introduced which enable one to separate applicability
domains for different approximations. Finally, a necessary condition for
the sum of thus calculated contributions to be correct is its independence
from any scale separating parameter.

In the present paper we successively pursue this line of reasoning for a
recalculation of the order $(\Z)^6m^2/M$ correction to hydrogen energy
levels. We discuss only $S$ levels since for higher angular momenta levels
the result is actually firmly established \cite{GKMY,I}. As far as the
controversy mentioned above concerns details of a regularization at the
subatomic scale, the result's dependence on a principal quantum number $n$
is also known. That's why we perform all the calculations for the ground
state and then restore the $n$ dependence in the final result.

To make the presentation self-contained we rederive some known results,
using sometimes new approaches. In Sec.2 the general outline of the problem
is given. Sections 3, 4 and 5 are devoted to the Coulomb, magnetic, and
seagull contributions, respectively. The correspondence between various
results is discussed in Conclusion. In Appendixes, we address a couple of
minor computational issues.

Throughout the paper the Coulomb gauge of electromagnetic potentials and
relativistic units $\hbar=c=1$ are used. Leaving aside the radiative
corrections we set $Z=1$ in what follows.

\section{General Outline}

The first recoil correction to a bound state energy of the relativistic
electron in the Coulomb field is an average value of the non-local operator,
\cite{Br,Sh,Y,PG,Sh2}
\be\label{main}
\dE_{rec}=-\fr{1}{M}\int\fr{d\om}{2\pi i}\la\lb\p-\D(\om,\rp)\rb
            G\lb\rp,\r|E+\om\rb\lb\p-\D(\om,\r)\rb\ra,
\ee
taken over an eigenstate of the Dirac equation in the Coulomb field,
\be
H\psi(\r) = E\psi(\r), \;\; H = \val\p +\beta m - \fr{\alpha}{r}.
\ee
In (\ref{main}), \p\ is the electron momentum operator, $\D(\om,\r)$
describes an exchange by the transverse (magnetic) quantum,
\be
\D(\om,\r)=\int\fr{d^3\k}{(2\pi)^3}\,e^{i\vec{k}\vec{r}}
\fr{4\pi\al\val_k }{k^2 - \om^2}, \;\;\;\;
\val_k \equiv \val - \fr{\k (\val\k)}{k^2},
\ee
while
\be
G\lb\rp,\r|E+\om\rb=\lb E+\om-\val\p -\beta m+\fr{\alpha}{r'}\rb^{-1}
\delta(\rp-\r)
\ee
is the Green's function for the Dirac equation in the Coulomb field. The
integration contour in (\ref{main}) goes from the minus infinity to zero
below the real axis, round zero from above and then proceeds to the plus
infinity above the real axis.

As far as we are going to calculate the correction (\ref{main})
perturbatively, i.e. as a power series in \al, it proves convenient to
decompose (\ref{main}) into three parts,
\be
\dE_{rec} = \C + \M + \Sg,
\ee
namely the Coulomb, magnetic and seagull contributions, corresponding to
$\p\p$, $\p\D +\D\p$ and $\D\D$ terms from (\ref{main}) respectively.

\section{Coulomb Contribution}

It is natural to continuously transform the integration contour into the sum
of two sub-contours, thus splitting the Coulomb contribution into two terms,
\be\label{C}
\C = \la \fr{p^2}{2M} \ra - \fr{1}{M} \la \p \Lambda_- \p \ra,
\ee
where $\Lambda_-$ is the projector to the set of negative-energy
Dirac-Coulomb eigenstates. The former term in (\ref{C}) results from the
integration along the upper half of the infinite circumference and its value
is determined by the atomic scale $p\sim m\al$. Being the average of the
local operator, this term can be easily calculated exactly. The latter term
in (\ref{C}) arises as an integral along the contour $C_-$, wrapping the
half-axis $(-\infty,0)$ in the counterclockwise direction. To the order we
discuss, this term is completely saturated by momenta from the relativistic
scale $p\sim m$. That's why it can be calculated without any regularization
\cite{PG,I}:
\be
- \fr{1}{M} \la \p \Lambda_- \p \ra_{\alpha^6} = \fr{m^2\al^6}{M}.
\ee

\section{Magnetic Contribution}

Using the identity
\be\label{id}
\la \p G \D + \D G \p \ra =
\fr{1}{\om}\la [\p,H] G \D + \D G [H,\p] + \{\p,\D\} \ra,
\ee
which follows directly from the equation for the Green's function,
we can extract from the general expression for the magnetic contribution,
\be\label{M}
\M = \fr{1}{M} \int \fr{d\om}{2\pi i} \la \p G \D + \D G \p \ra,
\ee
its local part,
\be\label{Inst}
\fr{1}{M}\int_{C_-}\fr{d\om}{2\pi i}\fr{1}{\om}
\la\left\{\p,\D(\om,\r)\right\}\ra=
-\fr{1}{2M} \la \left\{\p,\D(0,\r)\right\} \ra.
\ee
Due to the rapid convergency of the integral in (\ref{M}) at the infinity,
the integration contour can be reduced to $C_-$. By virtue of the virial
relations (see \cite{Eps} and references therein), the sum of local parts
of the Coulomb and magnetic contributions takes a simple form \cite{Sh}:
\be\label{ap}
\la \fr{p^2}{2M}-\fr{1}{2M} \left\{\p,\D(0,\r)\right\}  \ra
=\fr{m^2-E^2}{2M}.
\ee
Physically, this contribution to the recoil correction is induced by an
instantaneous part of the electron-nucleus interaction.

\subsection{Long Distances}

Immediate integration with respect to \om\ in (\ref{M}) gives \cite{I}:
\be\label{Mld}
\M = -\fr{\al}{M}\int\fr{d^3\k}{(2\pi)^3} \la \p \lb \sum_+
\fr{|m\rangle\langle m|}{k+E_m-E} - \sum_-\fr{|m\rangle\langle m|}{E-E_m+k}
\rb \fr{4\pi \val_k}{k}e^{i\vec{k}\vec{r}} \ra,
\ee
where $\sum_{+(-)}$ stands for the sum over discrete levels supplied by the
integral over positive- (negative-) energy part of the continuous spectrum.

\subsubsection{Positive Energies}

In the leading nonrelativistic approximation, the first term in
Eq.(\ref{Mld}) reads,
\be\label{M+ld}
\M_+=\fr{\al}{Mm}\int\fr{d^3\k}{(2\pi)^3}\la\p\, \G\lb\rp,\r|E-k\rb\fr{4\pi
      e^{i\vec{k}\vec{r}}}{k}\p_k\ra,
\ee
where $\G\lb\rp,\r|E-k\rb$ is the Green's function for the Schr\"odinger
equation in the Coulomb field, and the average is taken now over the
nonrelativistic wavefunction. For the ground state, we work with
\be
\psi(\r)=\sqrt{\fr{(m\alpha)^3}{\pi}}e^{-m\alpha r},\;\;
E=-\fr{m\alpha^2}{2}.
\ee
Only $p$-wave term from the partial expansion,
\be
\G\lb\rp,\r|\om\rb = \sum_l (-)^l(2l+1)P_l(\vec{n}'\vec{n})\G_l(r',r|\om),
\ee
survives the integration over the angles:
\be\label{M+}
\M_+=-\fr{m\al^3}{M\pi}\int_0^{\infty}dk\,k
\int_{-1}^1dx(1-x^2)\la \G_1\lb r',r|E-k\rb e^{ikrx}\ra.
\ee
For the nonrelativistic Green's function in the Coulomb field we use the
integral representation from the paper \cite{MS},
\be
\G_1\lb r',r\left|-\fr{\ka^2}{2m}\rb\right.=\fr{im}{2\pi\sqrt{r'r}}
\int_0^{\pi}\fr{ds}{\sin s}\fr{\exp i\left\{2\fr{m\alpha}{\kappa}s+
\fr{\kappa(r'+r)}{\tan s}\right\}}{1-e^{2i\fr{m\alpha}{\kappa}\pi}}
J_3\lb\fr{2\ka\sqrt{r'r}}{\sin s}\rb.
\ee
The integrals over $r$ and $r'$ in (\ref{M+}) are easily calculated after
expanding the Bessel function into the power series. The result can be
expressed in the form,
\be\label{MC}
\M_+=\fr{2^73m^5\al^6}{M\pi}\int_0^{\infty}
\fr{dk\;k}{\kappa^5}\int_{-1}^1dx(1-x^2)
\int_C dt\fr{t^{1-m\alpha/\kappa}}{(a-bt)^4}
\fr{1}{1-e^{2\pi i\fr{m\alpha}{\kappa}}},
\ee
where $\ka=\sqrt{2m(k-E)}$, the contour $C$ is the unit circumference
$|t|=1$ directed clockwise, and
\[
a=\lb 1+\fr{m\al}{\ka}\rb\lb 1+\fr{m\al}{\ka}-\fr{ikx}{\ka}\rb,\;\;\;
b=\lb 1-\fr{m\al}{\ka}\rb\lb 1-\fr{m\al}{\ka}+\fr{ikx}{\ka}\rb.
\]
Integration by parts conveniently extracts from the last integral in
(\ref{MC}) the terms non-vanishing at large momenta:
\be\label{M+red}
\M_+=-\fr{2^5m^2\al^5}{M\pi}\int_0^{1}dy(1-y^2)\int_{-1}^1dx(1-x^2)F(x,y),
\ee
where
\be\label{F}
F(x,y)=\fr{2}{b(a-b)^3}-\fr{1-y}{b^2(a-b)^2}-\fr{y(1-y)}{ab^2(a-b)}+
       \fr{1-y^2}{a^2b^2}-\fr{y(1-y^2)}{a^3b}
       \int_0^1\fr{dt\;t^{-y}}{1-\fr{b}{a}t},
\ee
and the new integration variable $y\equiv m\al/\ka$ is introduced. Since
\be
\fr{k}{\ka}=\fr{\al}{2}\fr{1-y^2}{y},
\ee
to get a power series expansion of (\ref{M+red}) with respect to \al\ up to
the
first order, we need an expansion of the integrand with respect to $y$ also
up to the first order (note that $a-b=4y-2ikx/\ka$):
\be\label{expan}
(1-y^2)F(x,y) \approx  \fr{2}{(a-b)^3} -\fr{1}{2(a-b)}+\fr{1}{2}-
            \fr{y^2}{2(a-b)} - \fr{y}{2} + y\ln (a-b).
\ee
Here the last term emerges as a result of expansion of the integral in
(\ref{F}),
\be
\int_0^1\fr{dt\;t^{-y}}{1-\fr{b}{a}t}=
\fr{1}{1-y}F\lb 1,1-y;2-y;\fr{b}{a}\rb,
\ee
where $F(1,1-y;2-y;b/a)$ is the Gauss hypergeometric function. Integrating
now (\ref{expan}) first with respect to $x$, and then with respect to $y$
from 0 to some $y_0$ ($\al^{1/2}\ll y_0 \ll 1$), we obtain,
\ba
\int_0^{y_0}dy(1-y^2)\int_{-1}^1dx(1-x^2)F(x,y) \approx
\fr{\pi}{32\alpha}-\fr{1}{48y_0^2}-\fr{1}{12}\ln\fr{4y_0^2}{\alpha}
+\fr{1}{48} &&\\ \non
 -\fr{1}{9}+\fr{2y_0}{3}+\fr{2y_0^2}{3}\ln 4y_0
-\fr{3y_0^2}{4}-\fr{\pi\alpha}{32}.&&
\ea
On the other hand, we can neglect \al\ in $F(x,y)$ on the interval
$[y_0,1]$. In the sum of two integrals, the dependence on the auxiliary
parameter $y_0$ disappears, and we come to the result,
\be\label{M+res}
\M_+=\fr{m^2\al^5}{M\pi}\left\{-\fr{\pi}{\al}+\fr{8}{3}\ln\fr{1}{\al}
+\fr{8}{3}\ln\fr{\mbox{Ry}}{\la E\ra_{1S}}+\fr{16}{3}\ln 2+\fr{32}{9}
-\pi\al\right\}.
\ee
Here the Bethe logarithm is introduced \cite{Gra},
\be
16\int_0^1 dy\,y
\fr{F\lb 1,1-y;2-y;\lb\fr{1-y}{1+y}\rb^2\rb-1}{(1+y)^4(1-y)}
= \ln\fr{\mbox{Ry}}{\la E\ra_{1S}}+2\ln 2 + \fr{11}{6}.
\ee
In (\ref{M+res}), the order $\al^4$ term is just the lowest-order
contribution to (\ref{Inst}), the order $\al^5$ terms are in accord with the
result of Salpeter \cite{Salp}, while the order $\al^6$ term coincides with
the retardation correction, found in \cite{I},Eq.(14) by the different
method.

It can be easily seen that the order $\al^6$ contribution to the
positive-energy part of (\ref{Mld}) is exhausted by the sum of those to
(\ref{Inst}) and (\ref{M+res}). Actually, relativistic corrections are at
least of the $\al^2$ relative order. The effect of retardation reveals
itself starting from the $\al^5$ order (\ref{M+res}). Hence the relativistic
corrections to the retardation are at least of the $\al^7$ order.

\subsubsection{Negative Energies}

Virtual transitions into negative-energy states give rise to the second term
in (\ref{Mld}). In the leading nonrelativistic approximation, it equals
\cite{I},
\be\label{M-}
\M_-=\fr{\al^2}{4m^2M}\int\fr{d^3\k}{(2\pi)^3}\la\fr{4\pi }{k'^2}
\fr{4\pi \k^2_{k'}}{k^2} \ra,
\ee
where $\k_{k'}=\k-\kp(\k\kp)/k'^2$, $\kp = \pp-\p-\k$, \p\ and \pp\ being
the arguments of the wavefunction and its conjugate respectively. The
integral over $k$ diverges logarithmically (leading linear divergency
vanishes due to the numerator which at $k \ar \infty$ becomes transverse to
itself, and hence rises only like $k$, not $k^2$). To treat this divergency
we use the following formal trick \cite{I}: subtract from (\ref{M-}) the
same expression with $k'^2+\lambda^2$ substituted in place of $k'^2$. For
$\lambda \gg m\al$, the subtracted term is completely determined by a scale
much less than the atomic one, so that we will find that term below using a
relativistic approach.

The regularized version of (\ref{M-}) can be written in the form
\be\label{M^r-}
\M_--\M_-^r= -\fr{\al^2}{4m^2M}
            \la (p'_i-p_i) \int\fr{d^3\k}{(2\pi)^3}\fr{4\pi k_j}{k^2}
            \lb \delta_{ij}-\fr{k'_ik'_j}{k'^2}\rb \lb\fr{4\pi}{k'^2} -
            \fr{4\pi}{k'^2+\lambda^2} \rb  \ra.
\ee
In the coordinate representation, the integral above is
\be
\fr{in_j}{r^2}\lb \delta_{ij} - \fr{\partial_i\partial_j}{\lambda^2} \rb
\fr{1-e^{-\lambda r}}{r} = \fr{in_i}{r^2}\int_0^{-\lambda}d\sigma
\lb 1 - \fr{\sigma^2}{\lambda^2} \rb e^{\sigma r}.
\ee
After substitution into (\ref{M^r-}) it gives
\be
\M_--\M_-^r= -\fr{\al^2}{4m^2M} \la 4\pi\delta(\r)\int_0^{-\lambda}d\sigma
               \lb 1 - \fr{\sigma^2}{\lambda^2} \rb + \fr{1}{r^2}
               \int_0^{-\lambda}d\sigma \sigma \lb 1 -
               \fr{\sigma^2}{\lambda^2} \rb e^{\sigma r} \ra.
\ee
Finally, the result of trivial calculation of the average over the ground
state reads,
\be\label{M-ld}
\M_--\M_-^r=\fr{m^2\al^6}{M} \lb 2 \ln\fr{\varepsilon}{\al} - 1 \rb,
\ee
where $\varepsilon \equiv \lambda/2m$.

\subsection{Short Distances}

Since in the nonrelativistic approximation the subtracted term, $\M_-^r$, is
ultraviolet divergent, we have to calculate it beyond this approximation, i.
e. using a relativistic approach. It proves more convenient in this approach
to postpone the integration over \om\ to the last stage of calculation. As
we will see below, the reversed order of integration (first over space
variables, then over frequency) makes the calculations quite simple. The fee
for the technical advantage is that a regulator contribution is calculated
not only for the negative-, but for the positive-energy part of \M\ also.
Surely, the instantaneous contribution can be left aside, so that only two
first terms from the r.h.s. of (\ref{id}) are considered below.

For the subtracted term, we have the new expansion parameter, $m\al/\lambda$,
and hence the Coulomb interaction during the single magnetic exchange can be
treated perturbatively. The order $m\al^6/M$ contributions arise due to only
two first terms of the Green's function expansion in the Coulomb
interaction, $G^{(0)}$ and $G^{(1)}$. Let us begin with the second
contribution:
\be
\M_G^r=\fr{2}{M}\int_{C_-}\fr{d\om}{2\pi
i}\fr{1}{\om}\la[\p,H]G^{(1)}\D^r\ra.
\ee
Here
\[
\D^r=\int\fr{d^3\k}{(2\pi)^3}\,e^{i\vec{k}\vec{r}}
\fr{4\pi \al \val_k }{k^2 + \lambda^2 - \om^2},
\]
and we can neglect atomic momenta in comparison with $\lambda$ and $m$:
\be
\M_G^r=-\fr{\al^3\psi^2}{\pi M}\int_{C_-}\fr{d\om}{i\om}\la\fr{4\pi\pp}{p'^2}
   \;\fr{2m+\om+\val\p\,'}{p'^2-\Omega^2}\;\fr{4\pi}{q^2}\;
   \fr{\om+\val\p}{p^2-\Omega^2}\;\fr{4\pi\val_p}{p^2-\K^2}\ra.
\ee
The notations of \cite{I} are used: $\psi^2\equiv |\psi(0)|^2$, the angle
brackets denote here integrations over $\p$ and $\p\,'$ together with the
average over the spinor $u_{\alpha}=\delta_{\alpha 1}$; $\vec{q} =
\p\,'-\p$; and
\[
\K\equiv \sqrt{\om^2-\lambda^2}, \;\;\;\;\;
\Omega\equiv\sqrt{2m\om+\om^2}.
\]
The average over the spin degrees of freedom gives
\be
\la(2m+\om+\val\p\,')(\om+\val\p)\val_p\pp\ra=\om\pp_p^2=\om\pp_p\vec{q}.
\ee
Then, after transition to the coordinate representation we get
\be
\M_G^r=\fr{2\al^3\psi^2}{mM}\int_{C_-}\fr{d\om}{i\om}\int_0^{\infty}dr\lb
\partial_i
\fr{e^{i\Omega r}-1}{\Omega^2 r} \rb n_j \left[\lb \delta_{ij}
+ \fr{\partial_i\partial_j}{\Omega^2} \rb \fr{e^{i\Omega r}-1}{r} -
(\Omega \ar \K) \right].
\ee
The integration over $r$ is simple but lengthy. It results in
\be
\M_G^r=-\fr{\al^3\psi^2}{2mM}\int_{C_-}\fr{d\om}{i\om}\left\{\fr{\K}{\Omega}
-\fr{\K^2}{\Omega^2}\ln \lb 1+\fr{\Omega}{\K}\rb+(\Omega\leftrightarrow\K)+
2\ln \lb 1+\fr{\K}{\Omega}\rb\right\}.
\ee
Here the contour of integration goes counterclockwise around the cut
connecting points $-2m$ and $-\lambda$. According to the Feynman rules,
$\Omega=i|\Omega|$, while $\K=+(-)|\K|$ on the lower (upper) edge of this
cut. Since the integrand is regular at small \om, we can put $\lambda=0$
(recall that $\lambda \ll m$) and get
\be
\M_G^r=\fr{\al^3\psi^2}{mM}\int_0^1 dx \lb \fr{\sqrt{1-x}}{x^{3/2}}-
\fr{1}{x^2}
\arctan\sqrt{\fr{x}{1-x}} - \fr{1}{\sqrt{x(1-x)}} \rb =
-\fr{3}{2}\fr{\pi\al^3\psi^2}{mM}.
\ee

To calculate a contribution due to $G^{(0)}$ we have to account properly for
the wavefunction's short-distance behavior:
\ba
\M_{\psi}^r=-\fr{\al^3\psi^2}{\pi M}\int_{C_-}\fr{d\om}{i\om}\la
\lb\fr{4\pi\pp}{p'^2}\;\fr{2m+\om+\val\p\,'}{p'^2-\Omega^2}\;
\fr{4\pi\val_q}{q^2-\K^2}\right.\right. && \\ \left.
\left.+\fr{4\pi\val_{p'}}{p'^2-\K^2}\;\fr{\om+\val\pp}{p'^2-\Omega^2}
\;\fr{4\pi\vec{q}}{q^2}\rb\fr{2m+\val\p}{p^2}\;\fr{4\pi}{p^2}\ra .&&\non
\ea
Averaging over the spin part of the wavefunction, we obtain
\ba
\M_{\psi}^r=-\fr{\al^3\psi^2}{\pi M}\int\fr{d\om}{i}\la\lb\fr{4m}{\om}+1\rb
\fr{4\pi}{p'^2(p'^2-\Omega^2)}\;\fr{4\pi}{q^2-\K^2}\;\fr{4\pi\p_q^2}{p^4}
\right. && \\
\left.-\;\fr{4\pi}{(p'^2-\K^2)(p'^2-\Omega^2)}\;\fr{4\pi}{q^2}\;
\fr{4\pi\p_{p'}^2}{p^4}\ra .&&\non
\ea
Again, the six-dimensional integral over \p\ and \pp\ turns into a simple
integral over $r$ in the coordinate representation, and equals
\ba
\M_{\psi}^r=\fr{2\al^3\psi^2}{M}\int\fr{d\om}{i}\left\{\lb\fr{4m}{\om}+1\rb
\left[\fr{1}{\Omega^2}\ln \lb 1+\fr{\Omega}{\K}\rb+\fr{1}{\K^2}
\ln\lb 1+\fr{\K}{\Omega}\rb -\fr{1}{\Omega\K}\right]\right. && \\
\left. +\fr{1}{2m\om}
\ln\fr{\K}{\Omega}\right\} .&&\non
\ea
Finally, the integration along the same contour as above gives for
non-vanishing in the limit $\varepsilon\ar 0$ terms:
\be\label{Mpsi}
\M_{\psi}^r=\fr{m^2\al^6}{M}\lb \fr{2}{\varepsilon}
-\fr{32}{9\pi\sqrt{\varepsilon}}\int_0^{\infty}
\fr{d\theta}{\sqrt{\cosh\theta}} + 2\ln \fr{1}{\varepsilon}\rb.
\ee
We see that as expected the logarithmic in $\varepsilon$ term cancels the
corresponding one in (\ref{M-ld}). The more singular in $\varepsilon$ terms
can only be the result of the regularization procedure applied to the
positive-energy contribution (\ref{M+res}). As far as the latter is
non-singular at short distances, this procedure is actually unnecessary, i.
e. it can produce only positive powers of $m\al/\lambda$. An explicit
calculation can be found in Appendix A.

\subsection{Total Magnetic Contribution}

So, in the sum of all contributions due to a single magnetic exchange, any
dependence on the scale separating parameter $\varepsilon$ cancels away, and
we get
\be\label{Mtot}
\M_{\alpha^6}+\la \fr{p^2}{2M} \ra_{\alpha^6}
=
\fr{m^2\al^6}{M}\lb -1+2\ln\fr{1}{\al}-1-\fr{3}{2}\rb.
\ee
Here $-1$ in the r.h.s. is due to the (long-distance) effect of retardation
(see (\ref{M+res}) and \cite{I}, Eq.(14)),
$2\ln\fr{1}{\al}-1$ comes from the whole range of scales from $m\al$ to $m$,
while $-3/2$ is the short-distance contribution.

\section{Seagull Contribution}

\subsection{Long Distances}

Again, the best suited way to analyze the atomic scale contribution
is to begin from taking the integral with respect to \om. It proves that
in the order of interest, only positive-energy intermediate states
are to be considered \cite{I}:
\be\label{S+}
\Sg_+= \fr{\al^2}{2M}\int\fr{d^3\k}{(2\pi)^3}
           \la\fr{4\pi}{k'^2}\fr{2\pp_{k'}+i\vec{\sigma}
           \times\kp}{2m}\fr{4\pi}{k^2}\fr{2\p_k+i\vec{\sigma}
           \times\k}{2m} \ra.
\ee
A simple power counting shows that only bilinear in \k\ and \kp\ term gives
rise to the ultraviolet divergency. To regularize this divergency,
we subtract from the divergent term the regulator contribution, which at
large distances equals to
\be\label{Sr}
-\fr{\al^2}{4m^2M} \la\fr{4\pi\kp}{k'^2+\lambda'^2}\;
\fr{4\pi\k}{k^2+\lambda^2}\ra,
\ee
while $m\al \ll \lambda,\lambda' \ll m$. In the coordinate representation,
the regularized version of (\ref{S+}) is
\be
\Sg_+-\Sg_+^r=\fr{\al^2}{4m^2M} \la 2\p\fr{1}{r^2}\p + \fr{1}{r^4}
-\lb\nabla\fr{e^{-\lambda'r}}{r}\rb\lb\nabla\fr{e^{-\lambda r}}{r}\rb\ra.
\ee
The average over the ground state reads ($\varepsilon'=\lambda'/2m$):
\be\label{Sld}
\Sg_+-\Sg_+^r=\fr{m^2\al^6}{M}\left\{
2\fr{\varepsilon'^2+\varepsilon'\varepsilon+
\varepsilon^2}{\alpha(\varepsilon'+\varepsilon)}+1-
2\ln\fr{\varepsilon'+\varepsilon}{\alpha}+
\fr{2\varepsilon'\varepsilon}{(\varepsilon'+\varepsilon)^2}\right\}.
\ee
Here 1 appears due to the non-singular operator $\p r^{-2}\p$. The first
term in the curly brackets represents the regulator contribution to the
previous order. In Appendix B, an appearance of this term as a
short-distance contribution to the $m\al^5/M$ order is shown explicitly.  In
what follows we calculate the subtracted term, whose non-relativistic
version (\ref{Sr}) is ultraviolet divergent, in the framework of a
relativistic approach.

\subsection{Short Distances}

Just like in case of the single magnetic exchange, only two first terms of
the Green's function expansion in the Coulomb interaction contribute to the
$m^2\alpha^6/M$ order. For the $G^{(1)}$'s contribution we have,
\be
\Sg_G^r = \fr{\al^3\psi^2}{2\pi M}\int_{C_-}\fr{d\om}{i}
\la\fr{4\pi\val_{p'}}{p'^2-\K'^2}\;\fr{\om+\val\pp}{p'^2-\Omega^2}
\;\fr{4\pi}{q^2}\;\fr{\om+\val\p}{p^2-\Omega^2}
\;\fr{4\pi\val_p}{p^2-\K^2}\ra.
\ee
Calculation along the same lines as in the case of $\M_G^r$ gives the result
\be\label{SG}
\Sg_G^r=\fr{\pi\al^3\psi^2}{Mm}\lb 4\ln 2 -2 \rb,
\ee
which is non-singular in the limit $\lambda, \lambda'\ar 0$.

As for the contribution due to $G^{(0)}$, it can be extracted from
\be\label{Spsi}
\fr{\al^3\psi^2}{2\pi M}\int_{C_-}\fr{d\om}{i}
\la\fr{4\pi\val_{p'}}{p'^2-\K'^2}\;\fr{\om+\val\pp}{p'^2-\Omega^2}
\;\fr{4\pi\val_q}{q^2-\K^2}\;\fr{2m+\val\p}{(p^2+\gamma^2)^2}\;4\pi\ra
+ (\lambda\leftrightarrow\lambda'),
\ee
as a zeroth-order term of the Laurent series in $\gamma\equiv m\al$ (this
series begins with an order $1/\gamma$ term describing the seagull
contribution to $m^2\alpha^5/M$ order at short distances discussed in
Appendix B). The average over the spin part of the wavefunction is
\be\label{numpsi}
\la 2m\om\val_{p'}\val_{q}+\val_{p'}\val\pp\val_q\val\p\ra=
-\lb \om^2 + [p'^2-\Omega^2]\rb\lb 1+\fr{(\pp\vec{q})^2}{p'^2 q^2}\rb
+2\pp\vec{q}.
\ee
The term in the square brackets can be omitted. In fact, the corresponding
part of (\ref{Spsi}) does not depend on $m$ and hence (merely on dimensional
grounds) contribute to the
$m^2\alpha^5/M$ order only. Then, the first term gives the non-singular in
the limit $\lambda, \lambda'\ar 0$ contribution:
\be\label{om2}
-\om^2\lb 1+\fr{(\pp\vec{q})^2}{p'^2 q^2}\rb\; \ar \;
\fr{\pi\al^3\psi^2}{Mm}\lb 1- 4\ln 2 \rb.
\ee
Finally, analysis of the last term from (\ref{numpsi}) deserves more care
since here we have the infrared singularity. Being integrated over the
space variables, this term gives:
\be
2\pp\vec{q}\; \ar\; \fr{2\al^3\psi^2}{Mm}\int_{C_-}\fr{d\om}{i\om}
\left\{ f(\Omega,\K)-f(\K',\K) \right\},
\ee
where
\be
f(x,y)=\ln \lb 1+\fr{x}{y} \rb -\fr{xy}{(x+y)^2}
\ee
(recall that $\K'=\sqrt{\om^2-\lambda'^2}$). For $\varepsilon \ll 1$ we
obtain
\be\label{fOmK}
\fr{2\al^3\psi^2}{Mm}\int_{C_-}\fr{d\om}{i\om} f(\Omega,\K) =
\fr{\pi\al^3\psi^2}{Mm}\lb -2\ln\fr{1}{\varepsilon}+4\ln 2 -1 \rb.
\ee
Calculation of the integral with $f(\K',\K)$ is a bit more cumbersome since
it does not contain a small parameter. The contour $C_-$ for this integral
encompasses in the counterclockwise direction the cut connecting the points
$-\lambda$ and $-\lambda'$. Continuous deformation of $C_-$ leads to the
following equation:
\be
\int_{C_-}d\om\ldots = \int_{C_+}d\om\ldots - 2\pi i\Res\limits_{\om=0}\ldots
-2\pi i\Res\limits_{\om=\infty}\ldots,
\ee
where $\ldots$ stands for $f(\K',\K)/\om$, and the
contour $C_+$ goes in the clockwise direction around the cut connecting the
points $\lambda$ and $\lambda'$. Using the evident relations,
\ba
\int_{C_+}d\om\ldots &=& - \int_{C_-}d\om\ldots\; ,\\
\Res\limits_{\om=0}\fr{1}{\om}f(\K',\K)&=&f(\lambda',\lambda)\; ,\\
\Res\limits_{\om=\infty}\fr{1}{\om}f(\K',\K)&=&-\ln 2 + \fr{1}{4}\; ,
\ea
we come to the result
\be\label{fKK}
\fr{2\al^3\psi^2}{Mm}\int_{C_-}\fr{d\om}{i\om}f(\K',\K)=
\fr{\pi\al^3\psi^2}{Mm}\lb 2\ln\fr{2\varepsilon}{\varepsilon+\varepsilon'} +
\fr{2\varepsilon\varepsilon'}{(\varepsilon+\varepsilon')^2}-\fr{1}{2} \rb.
\ee

\subsection{Total Seagull Contribution}

As can be seen from (\ref{Sld}), (\ref{SG}), (\ref{om2}), (\ref{fOmK}), and
(\ref{fKK}), the total seagull contribution to the $m^2\alpha^6/M$ order
does not depend on the scale separating parameters $\lambda$ and $\lambda'$,
and equals
\be\label{Stot}
\Sg_{\alpha^6}=\fr{m^2\al^6}{M}\lb 1-2\ln\fr{2}{\al}+\fr{1}{2}+4\ln 2-2\rb,
\ee
where 1 comes from the long distances, $4\ln 2-2$ from the short ones, while
the remaining terms gain their values on the whole range of scales, from
$m\al$ to $m$.

\section{Conclusions}

In complete accord with the result of Ref.\cite{FKMY}, the total correction
of the $m^2\alpha^6/M$ order does not contain $\ln\al$. It consists of two
terms,
\be\label{tot}
\dE_{rec} = \left.\fr{m^2-E^2}{2M}\right|_{\alpha^6}+
     \fr{m^2\al^6}{Mn^3}\lb 2\ln 2 - 3 \rb.
\ee
The former term is completely determined by the atomic scale and
depends non-trivially on a principal quantum number $n$,
\be
\left.\fr{m^2-E^2}{2M}\right|_{\alpha^6}= \fr{m^2\al^6}{2Mn^3}
\lb \fr{1}{4} + \fr{3}{4n} - \fr{2}{n^2} + \fr{1}{n^3} \rb.
\ee
As for the latter one, our calculations show that it has its origin
at the scale of the order of $m$.

The correction (\ref{tot}) shifts the hydrogen ground state by $-16.4$ kHz
and $2S$ state by $-1.9$ kHz. These figures are well comparable with the
uncertainties of the recent Lamb shift measurements \cite{exp}.

The result (\ref{tot}) differs from those obtained in Ref.\cite{PG,EG} and
in Ref.\cite{I}. Let us first discuss the origin of the difference in the
latter case. In \cite{I}, it was erroneously assumed that the cancellation
of singular operators at the atomic scale does not leave a non-vanishing
remainder. The present calculation shows that due to a difference in details
of a cut-off procedure used to regularize the average values of singular
operators, some finite contributions do survive the cancellation
process.

Unfortunately, the same error was repeated in Ref.\cite{EG}. The
long-distance contribution was found there in the framework of some
particular regularization scheme. Then it was added to the short-distance
contribution calculated in Refs.\cite{PG,I} by completely different
regularization procedures. The regularization dependence of the results
obtained in \cite{EG} can be seen, for example, in Eq.(29) of Ref.\cite{EG},
where the integration over $k'$ being limited above by a parameter $\sigma'$
gives rise to a finite (depending on $\sigma'/\sigma$) contribution to the
result.

The error made in Ref.\cite{PG} is a computational one. It is caused by
inaccurate treatment of the frequency dependence in the integral (42) of
Ref.\cite{PG} (ironically, by an evident typographical error, just the
important factors $(\om^2-k_1^2)^{-1}$ and $(\om^2-k_2^2)^{-1}$ are skipped
in Eq.(42) of Ref.\cite{PG}).  In what follows we rederive the result of the
present work employing the regularization scheme used by the authors of
Ref.\cite{PG}.

First of all, the result for the long-distance contribution (46) of
Ref.\cite{PG} ("the third term") is in accord with the result of the present
work ( 1 in (\ref{Sld})).

As for the remaining contributions, let us begin with one general note. In
their analysis of the integral (42), the authors of Ref.\cite{PG} use the
symmetrization in \om, since, as they wrote, "generally there are three
regions of photon energy $\om\sim\al^2$, $\om\sim\al$, and $\om\sim 1$ that
give a contribution and this middle region is almost eliminated by the
symmetrization". In order not to discuss here whether the middle region is
eliminated or not, we would like to recalculate the contributions of the
first and the second terms in (43) of Ref.\cite{PG} without the
symmetrization in \om. As far as the symmetrization procedure is no more
than a technical trick, a result of calculation should not depend on whether
this procedure is applied or not.

To get the high energy part of the first and second term contribution, we
put $\varepsilon'=\varepsilon=0$ in (\ref{Spsi}) and cut off the low energy
end $|\om|<m\epsilon$ from the contour $C_-$. Then the result for the
short-distance (high energy) contribution to the integral (42) of
Ref.\cite{PG} can be obtained:
\be\label{42sd}
\dE=\fr{m^2\al^6}{M}2\ln\fr{\epsilon}{2}.
\ee
The sum of the order $m^2\al^6/M$ contributions to Eqs.(51), (54) and (57)
of Ref.\cite{PG} is two times smaller. An extra factor one half emerges there
due to the symmetrization in \om, since the contribution of the contour
$C_+$, wrapping the half-axis $(m\epsilon,\infty)$, vanishes.

Turn now to the low energies. Only the second term of Eq.(43),
Ref.\cite{PG},  contributes there. According to (42) and (43) of
Ref.\cite{PG}, this contribution (with typos corrected) is,
\ba\label{2ndterm}
\dE&=&\fr{\al^2}{Mm}\int_{C_L}\fr{d\om}{2\pi i}\int\fr{d^3\k_1}{(2\pi)^3}
     \int\fr{d^3\k_2}{(2\pi)^3}\int\fr{d^3\p}{(2\pi)^3}\\ \non
     &&\psi(\p+\k_1)
     \fr{4\pi\k_1}{k_1^2-\om^2}\;\fr{1}{2m\om-p^2}\;
     \fr{4\pi\k_2}{k_2^2-\om^2}\psi(\p+\k_2).
\ea
Here the contour $C_L$ goes from $-m\epsilon$ to 0 below and then from 0 to
$m\epsilon$ above the real axis.  Recall now that the high energy
contribution (\ref{42sd}) is calculated on the assumption that
$\epsilon\gg\al$. It means that in (\ref{2ndterm}) we can neglect $p^2$
which is of the order of $(m\al)^2$, in comparison with $2m\om$ which will
be shown below to be of the order of $m^2\al$. Then we can easily come to
the coordinate representation and get,
\be
\dE= \fr{\al^2}{2Mm^2}\int_{C_L}\fr{d\om}{2\pi i}\;\fr{1}{\om-0}\;
     \la\lb\nabla\fr{e^{i|\omega|r}}{r}\rb^2 - \fr{1}{r^4} \ra.
\ee
Since the integration contour does not wrap the zero point, we can safely
add the operator $-1/r^4$ which is annihilated by the \om\
integration. The result of taking the average over the ground state is
\be
\dE=-\fr{2m^2\al^6}{M}\int_{C_L}\fr{d\om}{2\pi i}\;\fr{1}{\om-0}\;
\lb 2\ln\lb 1-\fr{i|\om|}{m\al}\rb +\fr{2i|\om|}{m\al}
\fr{1}{1-\fr{i|\omega|}{m\alpha}} + \fr{3}{2}\lb\fr{\om}{m\al}\rb^2
\fr{1}{1-\fr{i|\omega|}{m\alpha}} \rb.
\ee
Here we see that the natural scale for \om\ is in fact $m\al$. Since
$|\om|$ is positive on the lower half of $C_L$, the integral above written
in dimensionless units reads,
\be
\dE=-\fr{m^2\al^6}{\pi M}\int_0^{\epsilon/\alpha}dx \lb
\fr{4}{x}\mbox{arctan}x - \fr{4}{1+x^2} - 3\fr{x^2}{1+x^2} \rb.
\ee
The result of integration,
\be
\dE=\fr{m^2\al^6}{M}\lb 3\fr{\epsilon}{\pi\al} - 2\ln\fr{\epsilon}{\al}
    + \fr{1}{2} \rb,
\ee
being added to all the other seagull contributions, gives for the order
$m^2\al^6/M$ seagull correction:
\be
\Sg_{\alpha^6}=\fr{m^2\al^6}{M}\lb -2\ln\fr{1}{\al}+2\ln 2-\fr{1}{2}\rb,
\ee
in complete accord with the result (\ref{Stot}) of the present work.

\bigskip

{\bf Acknowledgments}\nopagebreak

The author is thankful to M.Eides, H.Grotch and A.Milstein for stimulating
discussions.  The work was supported by the Russian Foundation for
Basic Research, grant 97-02-18450, and by the program ``Universities
of Russia'', grant 95-0-5.5-130.

\bigskip

{\large \bf Appendix A}

Extra terms in (\ref{Mpsi}) should be canceled by the regulator counterpart
of (\ref{M+ld}) which differs from (\ref{M+ld}) by
$\sqrt{k^2+\lambda^2}$ placed instead of $k$. Just like in the main text we
approximate the sum over positive-energy intermediate states by the
nonrelativistic Green's function and the matrix element of \val\ by $\p/m$.
Within this approximation, the regulator contribution is
\be
\M_+^r=\fr{\al}{Mm}\int\fr{d^3\k}{(2\pi)^3}\la\p\,
      \G\lb\rp,\r|E-\sqrt{k^2+\lambda^2}\rb\fr{4\pi
      e^{i\vec{k}\vec{r}}}{\sqrt{k^2+\lambda^2}}\p_k\ra.
\ee
After the transformations, the regulator version of the expression
(\ref{MC}) is
\be
\M_+^r=\fr{2^73m^5\al^6}{M\pi}\int_0^{\infty}
\fr{dk\;k^2}{\kappa^5\om}\int_{-1}^1dx(1-x^2)
\int_C dt\fr{t^{1-m\alpha/\kappa}}{(a-bt)^4}
\fr{1}{1-e^{2\pi i\fr{m\alpha}{\kappa}}},
\ee
where $\ka=\sqrt{2m(\om-E)}$, $\om=\sqrt{k^2+\lambda^2}$, the contour $C$
and functions $a$ and $b$ are defined in the text. Only singular terms of
the expansion (\ref{expan}) operate at distances of the order of
$\lambda^{-1}$. For those terms the integrals over $k$ and $x$ become
elementary and give
\be
\M_+^r=\fr{m^2\al^6}{M}\left\{-
\fr{2}{\varepsilon^2}\lb \ln\fr{\varepsilon}{\al}-1 \rb+\fr{2}{\varepsilon}
-\fr{32}{9\pi\sqrt{\varepsilon}}\int_0^{\infty}
\fr{d\theta}{\sqrt{\cosh\theta}} \right\}.
\ee
So, the second and the third terms coincide with the corresponding terms in
(\ref{Mpsi}). The new singularity $\sim\varepsilon^{-2}$ is
nothing but the regulator contribution to the instantaneous part of the
magnetic exchange (\ref{Inst}):
\ba
\fr{1}{2M} \la \left\{\p,\D^r(0,\r)\right\} \ra
&\approx& 4\pi\al \la \fr{p'_i+p_i}{2m}\;\fr{p'_j+p_j}{2M}\;
\fr{\delta_{ij} -q_iq_j/q^2}{q^2+\lambda^2} \ra \\ \non
&=&2\fr{m^2\al^6}{M\varepsilon^2}\lb \ln\fr{\varepsilon}{\al}-1 \rb.
\ea

{\large \bf Appendix B}

The leading contribution to (\ref{Spsi}) is
\be
\Sg^r=\fr{8m^3\al^5}{M}\int_{C_-}\fr{d\om}{2\pi i}\int\fr{d^3\p}{(2\pi)^3}
      \fr{4\pi}{p^2-\K'^2}\;\fr{\om}{p^2-\Omega^2}\;\fr{4\pi}{p^2-\K^2}.
\ee
After the integration with respect to \p\ it turns into
\be
\Sg^r=\fr{m\al^5}{M\pi}\fr{1}{\varepsilon^2-\varepsilon'^2}\int_{C_-}
d\om\;\om\lb \fr{1}{\Omega+\K'} - \fr{1}{\Omega+\K}\rb.
\ee
Up to terms of the first order in $\varepsilon,\;\;\varepsilon'$, one gets
\be
\Sg^r=\fr{m^2\al^5}{M\pi}\lb 3 - 2\fr{\varepsilon^2\ln(2/\varepsilon)-
\varepsilon'^2\ln(2/\varepsilon')}{\varepsilon^2-\varepsilon'^2}-
2\pi\fr{\varepsilon'^2+\varepsilon'\varepsilon+
\varepsilon^2}{\varepsilon'+\varepsilon}\rb.
\ee
The last term compensates the leading contribution to (\ref{Sld}).

\end{document}